\input{aipcheck}

\documentclass[
    ,final            
  ]
  {aipproc}

\layoutstyle{6x9}
\newcommand{\NP}[1]{{ Nucl.\ Phys.\ } {\bf  #1}}

\newcommand{\PL}[1]{{ Phys.\ Lett.\ } {\bf  #1}}

\newcommand{\PRL}[1]{{ Phys.\ Rev.\ Lett.\ } {\bf  #1}}

\def\be{\begin{equation}}
\def\ee{\end{equation}}
\def\bea{\begin{eqnarray}}
\def\eea{\end{eqnarray}}

\newcommand{\lsim}{\raise.3ex\hbox{$<$\kern-.75em\lower1ex\hbox{$\sim$}}}
\newcommand{\ima}{{\mbox{Im}\,}}
\newcommand{\rea}{{\mbox{Re}\,}}

\begin{document}

\title{Light scalar and vector mesons in the large $N_c$ limit from
unitarized chiral perturbation theory}


\author{Jos\'e R. Pelaez}{
  address={Dept. F\'isica Te\'orica II. Universidad Complutense, 28040-Madrid, Spain}
}

\begin{abstract}
We review the large $N_c$ behavior of light scalar 
and vector resonances generated from unitarized meson-meson
scattering amplitudes at one loop in Chiral Perturbation Theory.
The vectors nicely follow the behavior expected for $\bar{q}q$ states,
whereas scalar mesons do not. This suggests that the main component
of light scalars is not $\bar{q}q$. We also comment on 
t-channel vector exchange as well as on the
large $N_c$ behavior of the mass splittings between the vectors
generated from the inverse amplitude method. 
\end{abstract}

\maketitle


\section{Introduction}
Chiral Perturbation Theory (ChPT) 
is the QCD low energy Effective Lagrangian
built as the most general derivative expansion
respecting SU(3) symmetry and 
containing only $\pi, K$ and $\eta$ mesons \cite{chpt1}. These particles 
are the QCD low energy degrees of freedom
since they are Goldstone bosons of the QCD spontaneous
chiral symmetry breaking.
For meson-meson scattering,  ChPT is an expansion in even
powers of momenta, $O(p^2), O(p^4)$...,
 over a scale $\Lambda_\chi\sim4\pi f_0\simeq 1\,$GeV.
Since the $u$, $d$ and $s$ quark
masses are so small compared with $\Lambda_\chi$ they
are introduced as perturbations, giving rise to the 
$\pi, K$ and $\eta$ masses, counted as $O(p^2)$. 
At each order, ChPT is the sum of {\it all terms}
 compatible with the symmetries,
multiplied by ``chiral  parameters'', that absorb
loop divergences order by order, yielding finite results.
The leading order is universal, containing only one parameter  $f_0$, 
that sets the scale of spontaneous symmetry breaking.
Different underlying dynamics correspond to different  values of the
higher order  parameters, called $L_i$, that, once renormalized,
depend on a regularization scale as 
$
 L_i(\mu_2)=L_i(\mu_1)+\Gamma_i\log(\mu_1/\mu_2)/{16\pi^2},
$
where $\Gamma_i$ are constants\cite{chpt1}.
In physical observables the $\mu$ dependence is canceled
with that of the loop integrals. 

\begin{table}[hbpt]
\caption{$O(p^4)$ chiral parameters ($\times10^{3}$) and their $N_c$ scaling.
In the ChPT column, $L_1$, $L_2$, $L_3$ come from \protect\cite{BijnensGasser} 
and  the rest from \protect\cite{chpt1}.
The IAM
columns correspond to different fits \protect\cite{GomezNicola:2001as}.
}
{\begin{tabular}{|c||c||c||c|c|c|}
\hline
$O(p^4)$&$N_c$&ChPT&IAM I&IAM II&IAM III\\
Parameter& Scaling &$\mu=770\,$MeV&$\mu=770\,$MeV&$\mu=770\,$MeV&$\mu=770\,$MeV\\
\hline
$L_1$
& $O(N_c)$
& $0.4\pm0.3$
& $0.56\pm0.10$ 
& $0.59\pm0.08$
& $0.60\pm0.09$
\\
$L_2$
& $O(N_c)$
& $1.35\pm0.3$ 
& $1.21\pm0.10$ 
& $1.18\pm0.10$
& $1.22\pm0.08$\\
$L_3 $  &
 $O(N_c)$&
 $-3.5\pm1.1$&
$-2.79\pm0.14$ 
&$-2.93\pm0.10$
& $-3.02\pm0.06$
\\
$L_4$
& $O(1)$
& $-0.3\pm0.5$& $-0.36\pm0.17$ 
& $0.2\pm0.004$
& 0 (fixed)\\
$L_5$
& $O(N_c)$
& $1.4\pm0.5$& $1.4\pm0.5$ 
& $1.8\pm0.08$
& $1.9\pm0.03$
\\
$L_6$
& $O(1)$
& $-0.2\pm0.3$& $0.07\pm0.08$ 
&$0\pm0.5$
&$-0.07\pm0.20$\\
$L_7 $  & $O(1)$ & 
$-0.4\pm0.2$&
$-0.44\pm0.15$ &
$-0.12\pm0.16$&
$-0.25\pm0.18$
\\
$L_8$
& $O(N_c)$
& $0.9\pm0.3$& $0.78\pm0.18$ 
&$0.78\pm0.7$
&$0.84\pm0.23$\\
\hline
$2L_1-L_2$
& $O(1)$
& $-0.55\pm0.7$& $0.09\pm0.10$ 
&$0.0\pm0.1$
&$-0.02\pm0.10$\\
\hline
\end{tabular}}
\end{table}

We will also make use of the large $N_c$ expansion \cite{'tHooft:1973jz},
which is the only
analytic approximation to QCD in the whole
energy region and provides
a clear definition of $\bar qq$ states, that become bound,
and whose
masses and widths behave as $O(1)$ and $O(1/N_c)$, respectively.
In fact, the $\pi,K,\eta$ masses scale as $O(1)$ and 
$f_0$ as $O(\sqrt{N_c})$.
The $L_i$ parameters that determine
meson-meson scattering up to $O(p^4)$ and their $N_c$ scaling
\cite{chpt1,chptlargen} 
is given in Table 1.
In order to apply the large $N_c$ expansion, the $\mu$ scale, a dependence suppressed by $1/N_c$,
has to be  chosen\cite{chpt1} between $\mu=$0.5 and 1 GeV.

In recent
years ChPT has been extended to higher energies by means of unitarization 
\cite{GomezNicola:2001as,Dobado:1996ps,Oller:1997ng,Pelaez:2003dy}. 
The main point is that the partial waves, $t_{IJ}$, of definite
angular momentum $J$ and isospin $I$,   in the elastic regime
satisfy the unitarity condition:
\begin{equation}
  \ima t_{IJ} =\sigma \vert t_{IJ}\vert^2, \quad \hbox{where} \; \sigma=\frac{2 q}{\sqrt{s}}\quad\Rightarrow \quad\ima \frac{1}{t_{IJ}}=-\sigma \quad\Rightarrow\quad
t_{IJ}=\frac{1}{\rea t_{IJ}^{-1} - i \sigma}, 
\end{equation}
where $q$ is the meson CM momentum. In order to have a unitary
amplitude we only need $\rea t^{-1}$, that can be obtained from ChPT: 
this is the
Inverse Amplitude Method (IAM) \cite{Dobado:1996ps,GomezNicola:2001as}.  
In this way, the IAM generates the $\rho$, $K^ *$, $\sigma$ and $\kappa$ resonances 
not initially present in ChPT, ensures unitarity
in the elastic region and respects the ChPT expansion.

When inelastic two-meson processes occur, the IAM can be generalized \cite{GomezNicola:2001as,Oller:1997ng} 
to $T\simeq(\rea T^{-1}-i \Sigma)^{-1}$, within a coupled channel formalism,
where 
$T$ is a matrix containing all partial waves 
between all physically accessible two-body states, whereas
$\Sigma$ is a diagonal
matrix with their corresponding phase spaces. 
Using one-loop ChPT calculations,
the coupled channel IAM provides  a remarkable description \cite{GomezNicola:2001as}
of  two-body $\pi$, K or $\eta$
scattering up to 1.2 GeV. In addition, it generates the $\rho$, $K^ *$, $\sigma$, $\kappa$,
$a_0(980)$,
$f_0(980)$ and the octet $\phi$. 
Such states are not included in the ChPT Lagrangian,
but each one has an associated pole
in the second Riemann sheet of its corresponding partial wave.
These poles 
appear already with the $L_i$ set used for standard ChPT,
which is compatible with the $L_i$ sets in Table 1, obtained from fits to data.
For narrow, Breit-Wigner like, resonances, their mass and width
is roughly given by  $\sqrt{s_{pole}}\sim M_R-i\,\Gamma_R/2$.
Furthermore, the IAM respects the $O(p^4)$
correct low energy expansion, with chiral parameters
compatible with standard ChPT.
Different IAM fits\cite{GomezNicola:2001as} 
are mostly due to different ChPT truncation schemes,
equivalent up to $O(p^ 4)$, and to the estimation of systematic errors in the data.

Since  ChPT amplitudes are 
renormalized, and therefore scale independent,
there are no  cutoffs or subtraction constants
where a spurious $N_c$ dependence could hide.
All the QCD $N_c$ dependence appears correctly
through the $L_i$, $f_0$ and the $\pi, K, \eta$ masses.

Recently\cite{Pelaez:2003dy}, 
by rescaling the ChPT parameters,  we have studied how
the resonances generated from unitarization behave in the large $N_c$ expansion,
around ``real life'', $N_c=3$.
Thus, in Fig.1 we see that
the modulus of partial waves associated to the $\rho(770)$ 
and $K^*(892)$ vector mesons presents a peak,
obtained from a fit to data,
that becomes narrower as $N_c$ increases, whereas the mass remains almost the same. This is exactly the behavior expected for a $\bar{q}q$ state, namely,
 $M\sim O(1)$, $\Gamma\sim O(1/N_c)$.

\begin{figure}
  \includegraphics[height=.115\textheight]{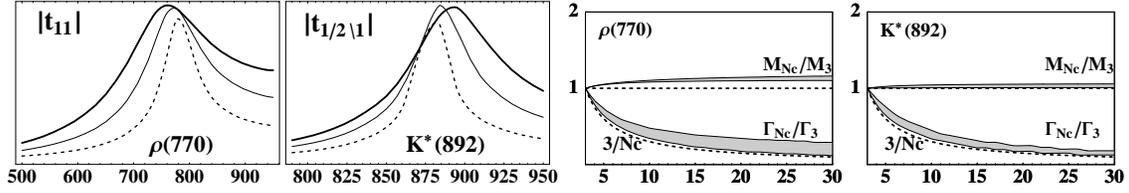}
  \caption{\footnotesize Left:
    Modulus of $\pi\pi$ and $\pi K$ elastic amplitudes versus
    $\sqrt{s}$ for $(I,J)=(1,1),(1/2,1)$: $N_c=3$ (thick line),
    $N_c=5$ (thin line) and $N_c=10$ (dotted line), scaled at
    $\mu=770\,$MeV.  Right: $\rho(770)$ and $K^*(892)$ pole positions:
    $\sqrt{s_{pole}}\equiv M-i\Gamma/2$ versus $N_c$. The gray areas
    cover the uncertainty $\mu=0.5-1\,$GeV. The dotted lines show the
    large $N_c$ scaling expected for a $\bar q q$ state. }
\end{figure}

\begin{figure}
  \includegraphics[height=.25\textheight]{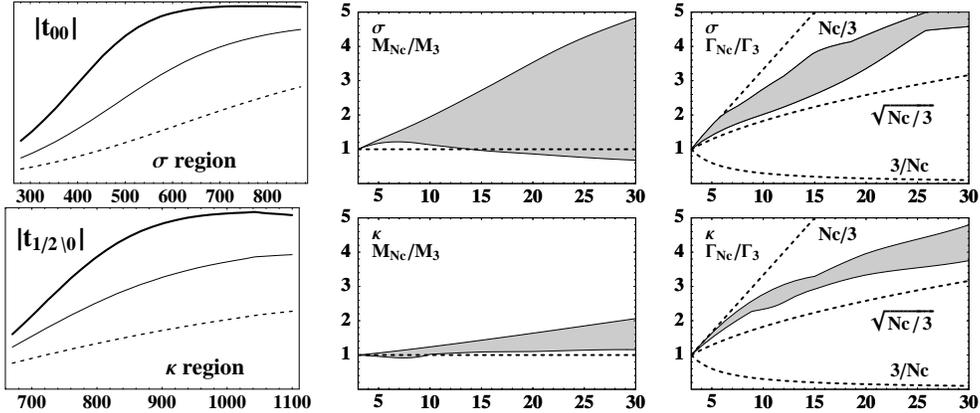}
  \caption{\footnotesize
(Top) Left:
Modulus of the $(I,J)=(0,0)$ scattering amplitude, versus $\sqrt{s}$
for $N_c=3$ (thick line), $N_c=5$ (thin line) and $N_c=10$ 
(dotted line), scaled at $\mu=770\,$MeV. Center: $N_c$ evolution
of the $\sigma$ mass. Right: $N_c$ evolution
of the $\sigma$ width. (Bottom): The same but for the $(1/2,0)$ amplitude
and the $\kappa$.
}
\end{figure}

In contrast, in Figure 2 we see the corresponding
behavior for the $\sigma$ (or $f_0(600)$) and the $\kappa$.
The results for the $f_0(980)$ and $a_0(980)$ 
(the latter, except in a corner of parameter space) are roughly similar, but 
more subtle \cite{Pelaez:2003dy}. 
It is evident that these {\it scalars behave completely different to $\bar{q}q$: The modulus of their
partial waves in the resonance region vanish and
their widths grow as $N_c$ increases from $N_c=3$}, 
as $O(N_c^{1/2})<\Gamma<O(N_c)$. These results have been confirmed later \cite{Uehara:2003ax}. Thus, we can conclude the following

\newcounter{input}
\begin{list}{$\bullet$}{
\setlength{\leftmargin}{0.2cm}
\setlength{\labelsep}{0.cm}}
\item {\it The \underline{dominant component}  of the $\sigma$ and $\kappa$ {\it in meson-meson scattering} does not behave as a $\bar{q}q$}.
\begin{list}{- }{
\setlength{\leftmargin}{0.2cm}
\setlength{\labelsep}{0.cm}}
  \item Why ``dominant''?
    Because, most likely, scalars are a mixture of different kind of
    states.  If the $\bar{q}q$ was {\it dominant}, they would behave
    as the $\rho$ or the $K^*$ in Figure 1.  {\it But it cannot be
      excluded that there is some smaller fraction of $\bar{q}q$.}
  \item Also, since scalars could 
be an admixture of states with different
nature and  wave functions, 
the small $\bar{q}q$ component could be concentrated in the core and better
seen in other reactions, whereas 
in scattering we are seeing mostly the outer region.
\end{list}
  \item {\it Two-meson and some tetraquark states \cite{Jaffe} have a consistent
      ``qualitative'' behavior}, i.e., both disappear in
    the continuum of the meson-meson scattering amplitude as $N_c$
    increases (also the glueballs for the $\sigma$ case, but not for the $\kappa$).  Waiting for more quantitative results, we have not been able
     to establish yet the nature of that dominant component, but
      two-meson states or some kind of tetraquarks are, qualitatively,
      candidates to form that dominant component.
\end{list}

Next we will address 
comments concerning the role of vector mesons in the IAM.

\section{t-channel vector meson exchange}

A common concern for people who have modeled meson-meson scattering
including explicitly resonance fields, is
the t-channel vector meson exchange, since it should
contribute sizably to $\pi\pi$ or $\pi K$ scattering.
Since the ChPT Lagrangian does not have an explicit $\rho$ or $K^*$ field,
and the usual resummations in the literature involve
only s-channel loops, it may seem that we are not taking into account
this contribution in the IAM. However,
the resonance saturation mechanism \cite{Ecker:1988te} explains
the $L_i$ values as the contact terms that remain once 
resonances heavier than pions, kaons and etas, are integrated
out from 
a chiral Lagrangian including all resonances.
In other words, resonance propagators are reduced to constants
when $M_V^2>>s$. From vector resonances
some $L_i$ get contributions of the form
\begin{equation}
L^V_i\sim 1/M_V^2,\label{eq:LV}
\end{equation}
where $M_V$ is the typical mass of the 
$\rho$
and $K^*$ multiplet.

Hence, a t-channel vector exchange, after integration of the heavy fields,
becomes a combination of $L_i$ parameters in the effective theory, and is 
thus {\it effectively}
included  in the
one loop amplitudes later used in the IAM \cite{Pelaez:2003dy}. 
This is another important reason to
consider the completely renormalized $O(p^4)$ ChPT amplitudes in the IAM,
because we can then use $L_i$ parameters compatible with standard ChPT, which
also ensure the correct \emph{low energy} expansion and the {\it effective}
crossed channel 
resonance exchange.

\section{ $\rho$ and $K^*$ mass splittings from the large $N_c$ IAM}

As shown in \cite{Pelaez:2003dy}, using $L_i$ parameters compatible with those
of standard ChPT, 
the IAM is able to fit the experimental data on many
meson-meson channels below 1.2 GeV, and in particular the different
physical masses of the
$\rho$ and $K^*$ resonances.
However, it has been remarked \cite{Leupold:2005ep} that
the vector  mass splittings are
\begin{equation}
  \label{eq:splittings}
  M_{K^*}^2=M_\rho^2+O(M_{\pi,K}^2)\simeq M_V^2+O(M_{\pi,K}^2)
\end{equation}
and that the $O(p^4)$ IAM does not yield this dependence in the large $N_c$
limit.
Certainly, it does not, but if we took into account the vector mass splittings,
 Eq.(\ref{eq:splittings}),
in the resonance saturation hypothesis,
 Eq.(\ref{eq:LV}),
we would find,
$$
L^{V+split}_i\sim 
\frac{1}{(M_V^2+O(M_{\pi,K}^2))}
\sim \frac{1}{M_V^2}(1+O(M_{\pi,K}^2))
\sim L_i^V\,(1+O(M_{\pi,K}^2)).
$$
But in ChPT, $O(M_{\pi,K}^2)$ counts as $O(p^2)$. Since $L^V_i$ is
multiplied by an $O(p^4)$ operator, the
splitting term contributes, at least, at $O(p^6)$.
Obviously, just with $O(p^4)$ ChPT
one does not have to get the $O(p^6)$ splitting terms. 
In particular, this does {\it not} mean that ``a systematic expansion in
powers of quark masses has not been performed'' \cite{Leupold:2005ep},
but that it has only been performed up to $O(p^4)$ in ChPT.
Therefore, \emph{nothing is to blame on the IAM,
but just on the truncation at $O(p^4)$}.
Any other unitarization scheme matching ChPT only up to $O(p^4)$ could also
fail at $O(p^6)$ in some observable.
In particular, even using $O(p^4)$  ChPT \emph{without unitarization}
to fit the low energy $\pi\pi$
and $\pi K$ scattering in the vector channels, the
resulting $L_i$ will not correspond to splittings
including $O(p^6)$ corrections.
Of course, when dealing with a truncated expansion to describe
data, the $L_i$ absorb the effect of higher orders, and
contain information on the physical splittings.

In summary, the observation  that the $M_{\pi,K}^2$ dependence of the
splittings is not obtained
at $O(p^4)$ is interesting, but the IAM
is not to blame for it, but just the $O(p^4)$ truncation. 
In particular, as we see in Fig.1,
it cannot be concluded  \cite{Leupold:2005ep} that
 the IAM at $O(p^4)$ {\it does not yield parametrically
reasonable large $N_c$ dependence for the vector \underline{masses}}, $O(1)$,
which is the 
only relevant issue for our discussion on the nature of resonances.
For mass {\it \underline{splittings}}, which is another issue,
one should consider the $O(p^6)$ {\it at least} \cite{inprep}.

\vspace*{-.1cm}
\begin{theacknowledgments}
The author thanks the HADRON05 organizers for creating such a nice 
working atmosphere during the Conference and S. Leupold and R. Kaminski
for comments and discussions. Work partially supported by the 
DGICYT contracts BFM2002-01868, and the EURIDICE network 
contract HPRN-CT-2002-00311 as well as the EU Hadron Physics Project, 
contract number RII3-CT-2004-506078.
\end{theacknowledgments}

\bibliographystyle{aipproc}   

\end{document}